____________________________________________________________________________________

# DARK ENERGY AND LIFE'S ULTIMATE FUTURE

**Rüdiger Vaas**

*Center for Philosophy and Foundations of Science, University of Gießen, Germany*
*Ruediger.Vaas@t-online.de*

____________________________________________________________________________________

Abstract: The discovery of the present accelerated expansion of space changed everything regarding cosmology and life's ultimate prospects. Both the optimistic scenarios of an ever (but decelerated) expanding universe and of a collapsing universe seem to be no longer available. The final future looks deadly dark. However, the fate of the universe and intelligence depends crucially on the nature of the still mysterious dark energy which drives the accelerated expansion. Depending on its – perhaps time-dependent – equation of state, there is a confusing number of different models now, popularly called Big Rip, Big Whimper, Big Decay, Big Crunch, Big Brunch, Big Splat, etc. This paper briefly reviews possibilities and problems. It also argues that even if our universe is finally doomed, perhaps that doesn't matter ultimately because there might be some kind of eternal recurrence.

## 1.     COSMIC CONSTRAINTS

The future is not what it used to be. There is a huge challenge and change going on in our understanding of the final fate of life and intelligence in our universe.[1] Even if all the very hard biological, psychological, sociological, and technological problems could be solved, which threaten life on Earth and perhaps also on other planets, and even if intelligent civilizations could succeed and colonize the galaxies, escaping the deaths of their parent stars and manage to use astronomical quantities of matter, energy, and information, the struggle is not won. If life and intelligence are to have a long-lasting future which is really worth to be called that way – and this ultimately means quasi-eternal existence and evolution –, it has to face the cosmological constraints.

We do not know very much about these constraints. But in the last few years everything that we thought we could know – and even would know soon –, got different. In 1970 Allan Sandage published a paper called „Cosmology: A Search for Two Numbers".[2] And until the 1990's it was thought that those two numbers would, in fact, predict the ultimate future of our universe. Since 1998 this has changed completely. Now we know those two numbers with a remarkable precision – but they turned out to be almost irrelevant after all concerning the very far future. Suddenly, a third quantity came in, which, as it seems now, rules everything.

The first of the two numbers which Sandage and his colleagues had been trying to determine for several decades is the **Hubble constant** $H_0$. It describes the expansion rate of our present universe. Its value is around 70 kilometers per second and megaparsec (a megaparsec is 3,26 million light-years). $H_o$ is not a true constant, but a time-dependent variable.

The second number is called the **deceleration parameter** $q_0$. It describes how fast $H_0$ changes in the future: $q_o = 1/2\ (\rho_{m,o}/\rho_{c,o}) - \Lambda c^2/3H_o^2$, where $\rho_{m,o}$ is the present mean density of matter (dark matter included), $\rho_{c,o} = 3H_o^2/8pG$ is the present critical density, which makes the universe flat (or Euclidean, thus there is no global curvature), and corresponds roughly to three hydrogen atoms per cubic meter, $\Lambda$ is the cosmological constant, c is the velocity of light, and G is Newton's gravitational constant.

For $q_o < 0$ the expansion gets faster (so in this case $q_o$ should more adequately called the acceleration parameter), for $q_o = 0$ the expansion rate is constant (what holds for an empty universe – mathematically simple, but not very interesting to us), for $q_o = 0,5$ the expansion will constantly decelerate and approach (but never quite reach) zero, and for $q_o > 0,5$ the expansion will stop and reverse someday.

The third mysterious quantity, which got everything mixed up, refers to what is now called **dark energy**.

The aim of this paper is to review the new situation and its confusing many scenarios, problems, and implications for the ultimate future of life. The discussion shall be restricted to the boundary conditions of fundamental physics and cosmology, and it shall survey the different possibilities only briefly without the physical details. Here it is assumed, at least for the sake of the argument, that no entirely new physical effects will change the whole picture (and even no variations of the fundamental constants of nature or that the topology of our universe is compact), nor that there are some kinds of nonphysical entities like Cartesian souls or transcendent Gods, which are beyond the reach of (current) science, but might alter the course of the universe in a non-predictable way. If we lived in a spiritual universe, or if idealism were true and matter was just a grand illusion, physical cosmology probably would be of no significance.

## 1.1     Dark energy

Different methodological approaches have recently led to a consistent and coherent picture of the universe we live in[3] – a picture which is sometimes called the **concordance model**, because it is the first world-view without obvious empirical or theoretical contradictions since the beginning of relativistic cosmology in 1917. According to this model, our universe emerged from a very hot and dense state, the Big Bang, 13,7 ± 0,2 billion years ago, expanded ever since and cooled down to 3 Kelvin yet. This is the temperature of the cosmic background radiation, which is left over from the primordial fireball and was released some 380000 years after the Big Bang, when the atoms were formed.

Many independent measurements of distant supernovae, the Hubble constant, the matter distribution and density, the large-scale structure of galaxies, quasar spectra, gravitational lensing effects, the Integrated Sachs Wolf effect, and the temperature fluctuations in the cosmic microwave background all point to a universe which has roughly the critical energy density $O_{total} = \rho_{m,o}/\rho_{c,o} + \Lambda c^2/3H_o^2 = 1$ and, therefore, is almost flat or Euclidean. The big surprise was that ordinary matter (about 4,4 ± 0,4 percent) and the still mysterious cold dark matter (23 ± 4 percent, probably unknown elementary particles without any electromagnetic interactions) together are adding up to only about a quarter of the total energy density. 73 ± 4 percent is made of what is now called dark energy. So if you imagine the universe as a cosmic cappuccino, the coffee stands for dark energy, the milk for dark matter, both of which we know almost nothing about; only the powdered chocolate would be what we are familiar with, namely ordinary matter made of protons, neutrons, electrons et cetera. This is the ironic success story of modern cosmology: Now we know what we don't know, and this is more than 95 percent of what the universe is made of.

Dark energy has negative pressure, therefore it is repulsive, thus it acts like antigravity (so, a better designation would be dark pressure, because it is the negative pressure which is abnormal – and not the energy density which remains positive). It is also what drives the accelerated expansion of space, which was discovered in 1998.[4] Space does not expand any slower due to the gravitational interaction of matter within, as it was long thought, but does expand faster! Measurements of distant supernovae indicate that this accelerated expansion started about five billion years ago.

But what about the future? What are the properties of dark energy, what physical effects will they have, and how will they affect the fate of life in the universe?

Whether space expands eternally (and how fast) or not depends on what is called the equation of state of dark energy. It is $p = \rho w$, where p stands for the pressure of the spatially homogeneous dark energy (and also for other, less exotic stuff), and $\rho$ for its energy density. Thus, the parameter w is the ratio between the pressure and the energy density.

Here are some numbers (not only for dark energy candidates):

- $w = 1/3$ represents electromagnetic radiation
- $w = 0$ represents non-relativistic matter
- $w = 1$ represents relativistic matter (for example within neutron stars)
- $w = -1$ represents the cosmological constant $\Lambda$, which was introduced by Albert Einstein in 1917 and wrongly withdrawn some years later
- $-1/3 > w > -1$ represents a scalar-field called kosmon or quintessence (here the value of w can change with time, for instance might decrease to almost $-1$ when the universe grows older)
- $w = -2/3$ represents topological defects (which would be primordial relics of the early universe, but probably do not exist in significant quantities within our horizon)
- $w < -1$ represents what is called phantom energy

It is not clear yet, whether the equation of state (and perhaps its time-derivative) describes the possibilities exhaustively and sufficiently, but at the moment these are the main alternatives. Thus, w can tell us something important about the nature of dark energy. And, assuming we are on the right track, dark energy determines the ultimate fate of the universe. So what are the options?

Before surveying them, a cautious comment seems to be appropriate.

## 1.2    Is dark energy real?

Accelerated expansion is only possible if at least one of the following assumptions is violated:

(1) The strong energy condition: the density and isotropic part of the pressure seen by all observers on timelike trajectories satisfy $\rho + 3p \geq 0$. – This is violated by dark energy.

(2) General Relativity as a large-scale description of our universe. – This is violated by modified theories of gravity, for example modified Friedmann equations, string theories with large extradimensions or relativistic versions of MOND (modified Newtonian dynamics).[5] By the way: Some theories of dark energy, some quintessence models for instance, violate not only (1) but, strictly speaking, also (2), implying modified Friedmann equations or new interactions.

(3) The assumed matter-dominated, homogeneous and isotropic Friedmann-Robertson-Walker cosmological model of the universe (even beyond our observational horizon). – This is violated if our whole observable universe is an underdense „bubble" within a denser environment.[6] This large-scale inhomogeneity might be the result of very long wavelength, super-horizon perturbations generated by a period of cosmic inflation in the early universe. The observed acceleration would be a „backreaction" from these perturbations without the assumption of new physics, that is the violation of (1) or (2). However, whether a violation of (3) alone is really sufficient to cause an accelerated expansion is still very controversial.[7] Another possibility for a violation of (3), or at least a wrong application of it, is this: Our Milky Way might be located at the center of an underdense „void" (with a radius of a few dozens or hundred light-years) in the large-scale distribution of galaxy clusters. This is improbable, but it would distort the measurement of $H_o$ nearby versus far away and could mimic an accelerated expansion.[8]

At present there is no convincing reason to drop dark energy, but one has to be aware of the alternatives. Only observations will tell which of the assumptions above is violated.

## 2. THE FUTURE OF LIFE AND THE UNIVERSE

### 2.1 The past future: chill or crunch

Before dark energy was discovered, there were only two possibilities in relativistic cosmology, depending on Sandage's famous two numbers (and $\Lambda = 0$):

(1) An infinite (either globally Euclidean or hyperbolic) universe is spatiotemporally open and will expand forever, but with an increasingly slower rate due to the decelerating effect of matter and energy. This scenario is called the **Big Whimper** (and sometimes the **Big Chill**), because everything finally fades away.

(2) A finite (but unlimited, that is spherically curved) universe is spatiotemporally closed and will start to contract after some time due to the gravitational interactions of its matter content, finally collapsing to a high density state called the **Big Crunch**. Even if the Big Crunch were to turn into a new Big Bang, nothing could survive this transition.

By the way, there is a scenario where in some sense the direction of time is switching at the maximum size of the finite universe when the expansion turns into contraction. While some have argued that even the psychological and thermodynamic arrow of time would run backwards (from the perspective of the expanding stage), and observers would still believe to live in an expanding phase, in a quantum cosmological framework everything with classical properties is destroyed in the maximum stage due to quantum interference, and the Big Bang and Big Crunch are ultimately the same, amusingly called the **Big Brunch**.[9] So in this scenario life has to cease even earlier than in the Big Crunch model, that is half way at the latest, so to speak.

Both scenarios challenge the far future of life tremendously, but there are at least some chances for long-term survival – provided, of course, that life can adapt itself to the changing conditions. It must learn to use the decreasing amount of energy and, perhaps, must even rebuild its own physical architecture (for example if protons were unstable and would decay). But if life is ultimately based on information-processing devices (that is on structure and organization, not substance), it might act like a software running on different kinds of hardware. And for this, many speculative options are conceivable. As John Desmond Bernal wrote already in 1929: „Finally, consciousness itself may end or vanish in a humanity that has become completely etheralized, losing the close-knit organism, becoming masses of atoms in space communicating by radiation, and ultimately perhaps resolving itself entirely into light. That may be an end or a beginning, but from here it is out of sight."

Frank Tipler[10] has argued that an advanced civilization can, in principle, colonize a closed universe entirely and should be able to manipulate its collapse to gain enough energy to live forever – forever with respect to subjective time (based on General Relativity's time dilatation) which is not to be confused with the finite objective time span such a contracting universe has.

Freeman Dyson[11] argued that in an open, eternally expanding universe with a decelerating expansion rate, life could also go on forever. „The pulse of life will beat more slowly as the temperature falls but will never stop".[12] Given longer and longer phases of hibernation, life-bearing devices could perform infinitely many calculations with a finite amount of energy. For a society of the size and complexity of our present civilization, for instance, $6 \cdot 10^{30}$ Joule would suffice – the amount of energy radiated away by our sun within just eight hours. It is controversial however, whether such a kind of existence is really eternal. First, because deleting old information to acquire new information costs energy, so perhaps only a finite number of thoughts could be instantiated. Second, it is unclear whether life is ultimately digital or analogous, that is, based on continuous processes. Only in the latter case a finite amount of energy is really sufficient and alarm clocks for hibernation wake up calls would not necessarily break. Lawrence Krauss and Glenn D. Starkman criticized Dyson's assumptions and believe that even in a decelerated eternal expansion of the universe life is ultimately doomed.[13] John Barrow and Sigbjørn Hervik argued, however, that arbitrarily weak anisotropies of the universe suffice to harness the temperature gradients created by gravitational tidal energy, and this should be enough to drive an information-processing machine arbitrarily long.[14]

Now with dark energy, the situation is not only more complicated, but also more desperate for the far future of life. But this depends crucially on the nature of dark energy which is not known yet.[15] Thus, many different alternatives have to be considered.

## 2.2 Resurrection and the accelerated future

The Big Whimper scenario is the most conservative or simple one. It is implied by the existence of a positive **cosmological constant** $\Lambda$ in the framework of relativistic cosmology. Here, the expansion goes on forever, and the expansion rate is approaching a final value $H_8 = v (\Lambda/3) = H_o v (1 - O_m)$ with $O_m = \rho_{m,o}/\rho_{c,o}$. Because of quantum effects at the horizon (analogous to Hawking radiation at the horizons of black holes) the universe cannot cool down to 0 Kelvin but reaches within a few hundred billion years a final temperature T of $T = 1/2pv (3/\Lambda) \approx 10^{-29}$ Kelvin (corresponding to $10^{-33}$ electron volts). This is the end for any living system because then it cannot radiate away waste heat – and there is no life without an energy gradient.[13]

The chances are better for some **quintessence** models. Here, things get very complicated for there are many different models. In most of them the accelerated expansion also lasts forever, but is slower than in the case of a cosmological constant. Depending on the models, perhaps the energy gradient remains.[16] Or the quintessence field decays, which might lead eventually to a decelerated eternal expansion[17] and a matter-dominated universe[18] or even to a revitalized universe if new matter is created out of the decaying field, and the origin and evolution of life would start again.[19] Or w could oscillate, causing alternating periods of accelerated and decelerated expansion.[20] Furthermore, it is not known whether quintessence is exactly homogeneous and whether it couples only gravitationally or in some other weak form, for example with neutrinos. Because the density of the dark energy ~ $(10^{-3}$ electron volts$)^4$ is in the order of the neutrino mass, it was argued that both are connected.[21] These ideas are mere speculations at the moment, but not ruled out yet.

So it seems to be very unlikely that a civilization can survive forever. In the end it will run out of energy. But this is a statistical argument, since it is based on the second law of thermodynamics. Entropy can decrease due to thermal **fluctuations**, and it is in principle possible that such fluctuations can sustain a civilization for an arbitrarily long time. The probability that a civilization will survive for some time is a sharply decreasing function of that time. However, for any finite time the probability is finite, and thus many civilizations in our universe will live longer than any given time, if the universe is large enough to allow those improbable events to occur. Of course, there is almost no chance that our successors are going to be among the lucky civilizations whose life will be prolonged by thermal fluctuations in such a lottery universe.

Furthermore, in an infinite future time might not be a problem. Eventually, anything could spontaneously pop into existence due to quantum fluctuations. They would mostly result in meaningless garbage, but a vanishingly small proportion shall be people, planets and parades of galaxies. This paper will reappear again, too. And such kinds of quantum resurrection might even spark a new Big Bang. According to Sean Carroll and Jennifer Chen one must be patient, however, and wait some $10^{10^{56}}$ years (if de Sitter vacuum is the „natural ground state").[22]

## 2.3 Big Splat and cosmic collapse

Despite the accelerated expansion today, the Big Crunch scenario is also not entirely refuted. However, as far as the latest measurements of w can tell, we are safe from the collapse for at least another 25 billion years – almost twice the age of our universe today.[23]

Within the framework of string cosmology, there is even a quite robust class of models, which describe our universe as a four-dimensional brane within a five-dimensional spacetime in which other 4D-branes might exist – literally parallel universes. And one of them might collide with our universe, perhaps countlessly often. Imagine two hands clapping repeatedly together. From the perspective within our universe every collision, sometimes called a **Big Splat**, looks like a Big Crunch and acts like a new Big Bang: The extradimension temporarily vanishes, and after the other brane-universe retreats, the expansion starts again and lasts for trillions of years. Note, that the branes are and remain infinite, only the inherent curvature looks like a collapse everywhere. One appeal of this Cyclic Universe[24] scenario is that it can explain the whole evolution by the action of dark energy. This is derived from string theory, that is it is the manifestation of a field called dilaton with has to do with the size of the extradimension. Whether a civilization could survive the Big Splat and spread again in the new expansion phase is not known. It seems to be unlikely, but perhaps advanced technology could create some kind of sheltering castles, for example with the help of black holes, to prevent the full brane collision locally. One could also envison a computer made of light, which would store all of our memories and information, might transmit it through the Big Splat, and recover it in the new expanding epoch of the universe. Whether this cyclic universe scenario or others are really future-enernal remains an open issue.
There are other possibilities for a Big Crunch, too.
Either dark energy leads to the contraction of our universe within a few dozens of billions of years only – a surprising possibility at least in the framework of supergravity and string theory. [25]
Or something like quintessence masks a negative cosmological constant today, which will dominate the far future if the accelerating cause vanishes. A negative cosmological constant leads inevitably – and independent of whether the universe is closed or not – to a collapse.
Or the current positive cosmological constant fluctuates due to quantum effects and, ultimately and irreversibly, turns negative. It was argued that this necessarily happens under quite weak conditions.[26] This would lead to a collapse of the universe within some trillions of years.
Whether a collapsing universe ruled by dark energy allows life to continue forever, as Frank Tipler has imagined[10], is unclear. If dark energy is a homogeneous property of spacetime geometry, it probably cannot be changed and accumulated somewhere like matter. But the manipulation of anisotropies are a necessary condition for a Tiplerian eschatology.

## 2.4      Big Rip and phantom energy

Dark energy offers a third option beyond the Big Whimper and Big Crunch scenarios if $w < -1$. Then the universe is ruled by what is called **phantom energy**.[27] It tears space literally apart and leads irrevocably to a complete disruption of everything, even atomic nuclei. This **Big Rip** could happen already soon in cosmological terms, that is in the order of tenths of billions of years. There is no mechanism by which life could stop such a process, so it seems to be doomed in this case. (By the way, there is even the possibility that some kind of matter ends in a Big Rip and some other kind does not[28], and there are also models without dark energy, but with some modifications of General Relativity containing the possibility for a Big Rip – or for a „Bigger Rip" due to an even more divergent scale factor within the next few billion years.[29] On the other hand, not all phantom energy scenarios end in a Big Rip singularity.[30])
One the other hand, phantom energy could be the solution for a problem of some oscillating universe scenarios with repeated Big Bang/Big Crunch cycles: Black holes might survive the crunch and grow in each cycle until they swallow the whole universe. But according to one model, phantom energy even tears black holes apart, effectively making them boil away.[31] In a string-theory-inspired braneworld model phantom energy disturbs fields in the fifth dimension outside our universe. Those fields might turn the parameter w of dark energy around, stop the Big Rip and let the universe recollapse. Although nothing would survive the Big Rip, new structures might form during the collapse. In addition, the higher dimensional fields would let the contraction bounce back to become the expansion of a new Big Bang.

## 2.5      A preposterous universe

So what is the ultimate future? At the moment we cannot say. Further measurements have to determine w and its time derivative as precise as possible. As it seems currently, w is close to –1 (that is $-0{,}8 > w > -1{,}2$, observationally speaking) and constant over time, which is consistent with a positive cosmological constant.[23] However, if $w \approx -1$, then we have an empirical limit we probably never can go beyond. This is a simple implication of the uncertainty and intrinsic error of every measurement. Thus, we can never empirically proof that $w = -1$, only that the error bars are very small around that value. Therefore, strictly speaking, the cosmological constant could only be established by other means, that is theory (and, as it was noted[26], Λ could even fluctuate nevertheless). The same is true for other limiting cases, for instance the widely hold flat or Euclidean universe with $O_{total} = 1$.

Even worse, we simply do not understand why the energy density of dark energy is about 0,7 today. From calculations in quantum physics it should be $10^{50}$ to $10^{120}$ times higher, which is completely unrealistic and the biggest discrepancy between theoretical prediction and observation ever in the history of physics. Also we do not understand why the dark energy density has roughly the same order as the matter density today – is this just pure chance or are there deeper connections?

So there is some kind of paradoxical situation: On the one hand we are in a golden age of cosmology, measuring the fundamental cosmological parameters with higher and higher precision – and on the other hand we do understand much less than expected. We seem to live in a „preposterous universe", as Michael Turner has said.

### 2.6    Big Decay or Big Hit

There is another depressing possibility which might even threaten us right now:

The **Big Decay**: The vacuum state in which we live in, that is our universe is in, might not be stable, but metastable.[32] Then it would be a kind of false vacuum like the one that, according to the inflationary scenario, might have driven the very early universe to an exponential (superluminal!) expansion until its field, the inflaton, had decayed (which might have released all the energy that turned into matter). Whether or not that epoch of inflation happened, and whether it is or is not related to the current accelerated (but much slower) expansion, for example as a kind of left-over, it is possible and even plausible that our universe has not reached its ground state (the „true vacuum"). But if the vacuum is metastable, it can and ultimately will decay. This could happen spontaneously, as a kind of quantum tunnel effect, or even by accident, for example due to an high-energy experiment of a very advanced civilization. From such a phase transition a wave of destruction would spread with nearly the velocity of light in every direction. It cannot be seen, and it almost instantly wipes out everything it hits without a warning. Similar effects would occur if there are very tiny extradimensions, as string theory claims, which are compactified, that is curled up, but could unfurl or decompactify. Even if only one tiny extradimension gets large, the whole universe would be burned to death. Perhaps new forms of life would arise out of the ashes, so to speak, but because the constants of nature would have changed, nothing could be like anything we know.

Not yet excluded but very improbable is the „rather unpleasant possibility" (Alexei Starobinsky) of a **Big Hit**: that our future world line will cross a space-time singularity, for example a gravitational shock wave with an infinite amplitude, or that it hits a finite-time singularity or a space-like curvature singularity which might form as a result of sudden growth of anisotropy and inhomogeneity at some moment during expansion or due to quantum gravitational effects.[33] (The Big Rip can be interpreted as a future singularity, too.)

### 2.7    Wormhole escapism and designer universes

All the scenarios reviewed here look more or less disappointing. If our universe is ultimately doomed, or at least the sufficient conditions for any possible information processing system disappear, the only chance for life would be to leave its universe and move to another place. There are bold speculations about **traversable wormholes** leading to other universes.[34] This seems to be possible at least in the framework of General Relativity (but like dark energy it violates some fundamental energy conditions). Perhaps wormholes could be found in nature and modified, or they could be built from scratch. If so, life could switch to another universe, escaping the death of its home. And if there is no life-friendly universe with the right conditions (physical constants and laws), an advanced civilization might even create a sort of replacement or rescue universe on its own. In fact, some renowned physicists have speculated about such a kind of world-making.[35] If such a switching of universes is possible, life might continue endlessly.

But even if our universe and every living being in it would be finally doomed, perhaps that doesn't matter ultimately. Because there could be infinitely many other universes and/or our universe might **recycle** itself due to new inflationary phase transitions out of black holes[36] or out of its high energy vacuum state, that is it creates new expanding bubbles growing to new universes elsewhere and cut their cords, metaphorically speaking.[37]

## 2.8 Eternal recurrence

Something strange is inevitable, if two conditions are true: Firstly, if our universe is infinite, or if there are infinitely many other universes with the same laws and constants. And, secondly, if quantum theory holds and there is, therefore, a finite number of possible states (that is due to Heisenberg's uncertainty relation there is no continuum of states and perhaps not even of space and time). If those two assumptions are valid, then according to Alexander Vilenkin every combination of discrete finite physical states are realized arbitrarily or infinitely often.[38] (Imagine a lattice built randomly out of zeros and ones: Every finite combination of zeros and ones, that is every local pattern occurs infinitely often.) Thus, there is a kind of spatial eternal recurrence.

This also implies that we would have perfect copies: **Doppelgänger** which are identical to us as far as quantum physics allows, and also Doppelgänger biographies, Doppelgänger earths, solar systems, Milky Ways and even Hubble volumes. Their distances are vast, but not infinite, and they could even be estimated, as Max Tegmark has shown: Our personal neighboring Doppelgänger should be $10^{10^{29}}$ meter apart, and another Doppelgänger Hubble volume, that is a region of space exactly like our observable universe, $10^{10^{115}}$ meter.[39] This means spatial eternal recurrence, but it could extend in time, which seems to be true either in a future-eternal inflationary or cyclic scenario with a flat universe. Thus, even if the history of our universe (and/or every universe) might lead to a global death, everything and every life-form might reappear over and over again, infinitely often both in space and time. Then, it is true that there is no personal eternal life, because every organism is doomed, but life as such could not be driven out of existence completely everywhere and everywhen. It would be truly eternal.

On the other hand, eternal recurrence seems to be absurd. And it is not only exact duplication – it is also every possible alternative, because all variations are equally real. As Alexander Vilenkin has said, some people „will be pleased to know that there are infinitely many […] regions where Al Gore is President and – yes – Elvis is still alive". Thus, physical potentiality and actuality would ultimately be the same. If so, the search for options and the struggle for life doesn't matter globally. Everything that might happen will happen at one place or another – in fact, it will happen infinitely often. This might be disappointing or encouraging. Perhaps this is only a matter of personal taste. However, it seems very strange and for many people even insulting, that we are not unique, and that everything we try might succeed here but not elsewhere and vice versa – infinitely often. However, as Steven Weinberg reminded us, „our mistake is not that we take our theories too seriously, but that we do not take them seriously enough".[40]

## 2.9 Conclusion

In conclusion, we can safely say that the future is not, what it used to be. There is a diverse range of alternative cosmological scenarios (both for the future and for the past[41]), some of which look really eery. It is premature, however, to announce the ultimate and unavoidable end of everything in the very far future. But undoubtly huge challenges are imminent. As Niels Bohr once joked, it is difficult to make predictions, especially about the future. But it would be of no surprise, if big surprises out there are still waiting for us to be discovered. Perhaps the universe, as John Burdon Sanderson Haldane once said, is „not only queerer than we imagined, it's queerer than we can imagine".

**Acknowledgments**

It is a pleasure to thank Hans-Joachim Blome, Rob Caldwell, Freeman Dyson, Gia Dvali, Katherine Freese, Gerson Goldhaber, Alan Guth, Renata Kallosh, Claus Kiefer, Robert Kirshner, Lawrence Krauss, John Leslie, Andrei Linde, Mario Livio, Saul Perlmutter, Wolfgang Priester, Adam Riess, Antonio Riotto, Subir Sarkar, Larry Schulman, Lee Smolin, Glenn Starkman, Paul Steinhardt, Michael Turner, Neil Turok, Carsten Van der Bruck, Alex Vilenkin, Christof Wetterich, and H. Dieter Zeh for comments and discussion, André Spiegel for his kind support, and Vladimir Burdyuzha as well as Claudius Gros for the invitation to „The Future of Life and the Future of Our Civilization" symposium at the Johann Wolfgang Goethe University in Frankfurt am Main, Germany, where an earlier version of this paper was presented on May, 5th 2005.

**REFERENCES**

1. F. C. Adams, G. Laughlin, A dying universe, *Rev. Mod. Phys.* **69:** 337–372 (1997); astro-ph/9701131. – F. C. Adams, G. Laughlin, *The Five Ages of the Universe* (Free Press, New York, 1999). – M. M. Cirkovic, A Resource Letter on Physical Eschatology, *Am. J. Physics* **71:** 122–133 (2003); astro-ph/0211413. – P. Davies, *The Last Three Minutes* (New York, Basic Books, 1994). – G. F. R. Ellis (ed.), *The Far-Future Universe* (Templeton Press, Radnor, 2002). – J. N. Islam, *The Ultimate Fate of the Universe* (Cambridge University Press, Cambridge, 1983). – A. Loeb, The Long-Term Future of Extragalactic Astronomy. *Phys. Rev.* **D65:** 047301 (2002); astro-ph/0107568. – N. Prantzos, *Our Cosmic Future* (Cambridge University Press, Cambridge, 2000 [1998]). – R. Vaas, Die fernste Zukunft, in: U. Anton: *Die Lebensboten* (Heyne, München, 2004), 255–320. – R. Vaas, Die ferne Zukunft des Lebens im All, in: S. Mamczak, W. Jeschke (eds.): *Das Science Fiction Jahr 2004* (Heyne, München, 2004), 512–594. – R. Vaas, Ein Universum nach Maß?, in: J. Hübner, I.-O. Stamatescu, D. Weber (eds.): *Theologie und Kosmologie* (Mohr Siebeck, Tübingen, 2004), 375–498.
2. A. Sandage, Cosmology: A Search for Two Numbers, *Phys. Today* **23:** 34–41 (1970).
3. D. N. Spergel et al., First Year Wilkinson Microwave Anisotropy Probe (WMAP) Observations: Determination of Cosmological Parameters, *Astrophys. J. Suppl.* **148:** 175-194 (2003).
4. S. Perlmutter et al., Measurements of Omega and Lambda from 42 High-Redshift Supernovae, *Astrophys. J.* **517:** 565–586 (1999); astro-ph/9812133. – A. G. Riess et al., Observational Evidence from Supernovae for an Accelerating Universe and a Cosmological Constant, *Astron. J.* **116:** 1009–1038 (1998); astro-ph/9805201. – S. Perlmutter, Supernovae, Dark Energy, and the Accelerating Universe, *Phys. Today* **4:** 53–60 (2003).
5. P. D. Mannheim, Alternatives to Dark Matter and Dark Energy; astro-ph/0505266. – S. M. Carroll et al., Is Cosmic Speed-Up Due to New Gravitational Physics?, *Phys. Rev.* **D70:** 043528 (2004); astro-ph/0306438.
6. E. W. Kolb, S. Matarrese, A. Riotto, On cosmic acceleration without dark energy; astro-ph/0506534. – E. W. Kolb et al., Primordial inflation explains why the universe is accelerating today; hep-th/0503117. – E. Barausse, S. Matarrese, A. Riotto, The Effect of Inhomogeneities on the Luminosity Distance-Redshift Relation: Is Dark Energy Necessary in a Perturbed Universe?, *Phys. Rev.* **D71:** 063537 (2005); astro-ph/0501152. – J. W. Moffat, Late-time Inhomogeneity and Acceleration Without Dark Energy; astro-ph/0505326. – D. L. Wiltshire, Viable inhomogeneous model universe without dark energy from primordial inflation; gr-qc/0503099.
7. C. M. Hirata, U. Seljak, Can superhorizon cosmological perturbations explain the acceleration of the universe?; astro-ph/0503582. – E. R. Siegel, J. N. Fry, Effects of Inhomogeneities on Cosmic Expansion; astro-ph/0504421. – E. E. Flanagan, Can superhorizon perturbations drive the acceleration of the Universe?, *Phys. Rev.* **D71:** 103521 (2005); hep-th/0503202.
8. K. Tomita: A Local Void and the Accelerating Universe, *Mon. Not. Roy. Astron. Soc.* **326:** 287–292 (2001); astro-ph/0011484. – A. Blanchard et al., An alternative to the cosmological ‚concordance model', *Astron. Astrophys.* **412:** 35–44 (2003); astro-ph/0304237.
9. C. Kiefer, H. D. Zeh, Arrow of time in a recollapsing quantum universe, *Phys. Rev.* **D51:** 4145–4153 (1995).
10. F. J. Tipler, *The Physics of Immortality* (Anchor Books, New York, 1994).
11. F. J. Dyson, Time without end, *Rev. Mod. Phys.* **51:** 447–460 (1979).
12. F. J. Dyson, *Infinite in all directions* (Penguin, London, 1990 [1985]), 111.
13. L. M. Krauss, G. D. Starkman, Life, The Universe, and Nothing, *Astrophys. J.* **531:** 22–30 (2000); astro-ph/9902189
14. J. D. Barrow, S. Hervik, Indefinite Information Processing in Ever-expanding Universes, *Phys. Lett.* **B566:** 1–7 (2003); gr-qc/0302076.

15. P. J. E. Peebles, B. Ratra, The Cosmological Constant and Dark Energy, *Rev. Mod. Phys.* **75:** 559–606 (2003); astro-ph/0207347. – R. P. Kirshner, Throwing Light on Dark Energy. *Science* **300:** 1914–1918 (2003). – J. C. N. de Araujo, The dark energy-dominated Universe, *Astropart. Phys.* **23:** 279–286 (2005); astro-ph/0503099.
16. K. Freese, Cardassian Expansion: Dark Energy Density from Modified Friedmann Equations. *New Astron. Rev.* **49:** 103–109 (2005); astro-ph/0501675.
17. U. Alam, V. Sahni, A. A. Starobinsky, Can dark energy be decaying?, *JCAP* **0304:** 002 (2003); astro-ph/0302302.
18. J. Barrow, R. Bean, J. Magueijo, Can the Universe escape eternal acceleration?, *Mon. Not. Roy. Astron. Soc.* **316:** L41–L44 (2000); astro-ph/0004321.
19. J. P. Ostriker, P. Steinhardt, The Quintessential Universe, *Sci. Am.* **284** (1), 47–53 (2001).
20. S. Dodelson, M. Kaplinghat, E. Stewart, Solving the Coincidence Problem: Tracking Oscillating Energy, *Phys. Rev. Lett.* **85:** 5276–5279 (2000); astro-ph/0002360. – K. Griest, Toward a Possible Solution to the Cosmic Coincidence Problem, *Phys. Rev.* **D66:** 123501 (2002); astro-ph/0202052.
21. R. D. Peccei, Neutrino Models of Dark Energy; hep-ph/0411137.
22. S. M. Carroll, J. Chen, Spontaneous Inflation and the Origin of the Arrow of Time; hep-th/0410270.
23. A. G. Riess et al., Type Ia Supernova Discoveries at z>1 From the Hubble Space Telescope: Evidence for Past Deceleration and Constraints on Dark Energy Evolution, *Astrophys. J.* **607:** 665–687 (2004); astro-ph/0402512.
24. P. J. Steinhardt, N. Turok, Cosmic Evolution in a Cyclic Universe, *Phys. Rev.* **D65:** 126003 (2002); hep-th/0111098. – P. J. Steinhardt, N. Turok: The Cyclic Model Simplified, *New Astron. Rev.* **49:** 43–57 (2005); astro-ph/0404480.
25. R. Kallosh, A. Linde, Dark Energy and the Fate of the Universe, *JCAP* **0302:** 002 (2003); astro-ph/0301087. – R. Kallosh et al., Observational Bounds on Cosmic Doomsday, *JCAP* **0310:** 015 (2003); astro-ph/0307185.
26. J. Garriga, A. Vilenkin, Testable anthropic predictions for dark energy, *Phys. Rev.* **D67:** 043503 (2003); astro-ph/0210358.
27. R. R. Caldwell, M. Kamionkowski, N. N. Weinberg, Phantom Energy and Cosmic Doomsday, *Phys. Rev. Lett.* **91:** 071301 (2003); astro-ph/0302506.
28. M. Gasperini, Towards a future singularity?, *Int. J. Mod. Phys.* **D13:** 2267–2274 (2004); gr-qc/0405083.
29. P. H. Frampton, T. Takahashi, Bigger Rip with No Dark Energy, *Astropart. Phys.* **22:** 307–312 (2004); astro-ph/0405333.
30. P. Wu, H. Yu, Avoidance of Big Rip In Phantom Cosmology by Gravitational Back Reaction; astro-ph/0407424; – R. Curbelo, T. Gonzalez, I. Quiros, Interacting Phantom Energy and Avoidance of the Big Rip Singularity; astro-ph/0502141.
31. M. G. Brown, K. Freese, W. H. Kinney, The Phantom Bounce: A New Oscillating Cosmology; astro-ph/0405353.
32. J. Leslie, *The End of the World* (Routledge, London and New York, 1998 [1996]), 108–122.
33. A. Starobinsky, Future and Origin of our Universe: Modern View, in: V. Burdyuzha, G. Khozin (eds), *The Future of the Universe and the Future of Our Civilization* (World Scientific, Singapore etc., 2000), 71–84; astro-ph/9912054. – J. D. Barrow, C. G. Tsagas, New Isotropic and Anisotropic Sudden Singularities, Class. Quant. Grav. **22:** 1563–1571 (2005); gr-qc/0411045.
34. M. Visser, *Lorentzian Wormholes* (American Institute of Physics Press, Woodbury, 1996). – R. Vaas, *Tunnel durch Raum und Zeit* (Franckh-Kosmos, Stuttgart, 2005).
35. E. Farhi, A. H. Guth, An obstacle to creating a universe in the laboratory, *Phys. Lett.* **B183:** 149–155 (1987). – V. P. Frolov, M. A. Markov, M. A. Mukhanov, Through a black hole into a new universe?, *Phys. Lett.* **B216:** 272–276 (1989). – A. Linde, Hard Art of the Universe Creation, *Nucl. Phys.* **B372:** 421–442 (1992); hep-th/9110037. – E. R. Harrison, The natural selection of universes containing intelligent life, *Quart. J. R. astr. Soc.* **36:** 193–203 (1995).
36. L. Smolin, Did the universe evolve? *Class. Quant. Grav.* **9:** 173–191 (1992). – R. Vaas, Is there a Darwinian Evolution of the Cosmos?; gr-qc/0205119.
37. K. M. Lee, E. J. Weinberg, Decay Of The True Vacuum In Curved Space-Time, *Phys. Rev.* **D36:** 1088–1094 (1987). – J. Garriga, A. Vilenkin, Recycling universe, *Phys. Rev.* **D57:** 2230–2244 (1998); astro-ph/9707292.
38. J. Garriga, A. Vilenkin, Many worlds in one, *Phys. Rev.* **D64:** 043511 (2001); gr-qc/0102010. – J. Knobe, K. D. Olum, A. Vilenkin, Philosophical Implications of Inflationary Cosmology; physics/0302071.
39. M. Tegmark, Parallel Universes, in: J. Barrow, P. C. W. Davies, C. L. Harper jr. (eds.): *Science and Ultimate Reality* (Cambridge University Press, Cambridge, 2004), 459–491; astro-ph/0302131.
40. S. Weinberg, *The First Three Minutes* (Basic Books, New York, 1977), 131.
41. R. Vaas, Time before Time; physics/0408111.